\documentclass[prl,aps,floatfix,twocolumn]{revtex4}%
\usepackage{epsfig}
\usepackage{amsmath}
\usepackage{amsfonts}
\usepackage{amssymb}
\usepackage{xcolor}
\usepackage[caption=false]{subfig}
\usepackage{graphicx}%
\providecommand{\U}[1]{\protect\rule{.1in}{.1in}}
\providecommand{\U}[1]{\protect\rule{.1in}{.1in}}

\begin{document}
\title{Anisotropy of magnetic interactions and symmetry of the order parameter in
unconventional superconductor Sr$_{2}$RuO$_{4}$}
\author{Bongjae Kim$^{1}$}
\thanks{These two authors contributed equally}
\author{Sergii Khmelevskyi$^{1,2}$\textsuperscript{\thefootnote}}
\author{Igor I. Mazin$^{3}$}
\author{Daniel F. Agterberg$^{4}$}
\author{Cesare Franchini$^{1}$}
\email{cesare.franchini@univie.ac.at}
\affiliation{$^{1}$ University of Vienna, Faculty of Physics and Center for Computational
Materials Science, Vienna A-1090, Austria}
\affiliation{$^{2}$ Center for Computational Materials Science, Institute for Applied
Physics, Vienna University of Technology, Wiedner Hauptstrasse $8$ - $10$,
$1040$ Vienna, Austria}
\affiliation{$^{3}$Code 6393, Naval Research Laboratory, Washington, D.C. 20375, USA}
\affiliation{$^{4}$ Department of Physics, University of Wisconsin, Milwaukee, Wisconsin
53201, USA}
\date[Dated: ]{\today}
\maketitle

\textbf{
Sr$_2$RuO$_4$ is the best candidate for spin-triplet superconductivity,
an unusual and elusive superconducting state of fundamental importance. In the
last three decades Sr$_2$RuO$_4$ has been very carefully studied and despite
its apparent simplicity when compared with strongly correlated high-$T_{c}$ cuprates,
for which the pairing symmetry is understood, there is no scenario that can explain
all the major experimental observations, a conundrum that has generated tremendous interest.
Here we present a density-functional based analysis of magnetic
interactions in Sr$_{2}$RuO$_{4}$ and discuss the role of magnetic anisotropy
in its unconventional superconductivity. Our goal is twofold. First, we access
the possibility of the superconducting order parameter rotation in an external
magnetic field of 200 Oe, and conclude that the spin-orbit interaction in this
material is several orders of magnitude too strong to be consistent with this
hypothesis. Thus, the observed invariance of the Knight shift across }%
$T$\textbf{$_{c}$ has no plausible explanation, and casts doubt on using the
Knight shift as an ultimate litmus paper for the pairing symmetry. Second, we
propose a quantitative double-exchange-like model for combining itinerant
fermions with an anisotropic Heisenberg magnetic Hamiltonian. This model is
complementary to the Hubbard-model-based calculations published so far, and
forms an alternative framework for exploring superconducting symmetry in
Sr$_{2}$RuO$_{4}.$ As an example, we use this model to analyze the degeneracy
between various }$p-$\textbf{triplet states in the simplest mean-field
approximation, and show that it splits into a single and two doublets with the
ground state defined by the competition between the \textquotedblleft
Ising\textquotedblright\ and \textquotedblleft compass\textquotedblright%
\ anisotropic terms.}

\section{Introduction}

Superconductivity in Sr$_{2}$RuO$_{4},$ even though it occurs at a rather low
temperature, has been attracting attention comparable to that attached to
high-temperature superconductors~\cite{review}. For many years the dominant
opinion was that it represents a unique example of a chiral triplet pairing
state~\cite{K,K-O,chi,msr}. Interestingly, the original premise that led to
this hypothesis was the presumed proximity of Sr$_{2}$RuO$_{4}$ to
ferromagnetism, and thus it was touted as a 3D analogue of $^{3}%
$He~\cite{Sigrist-Rice,MS1}. It was soon discovered, first
theoretically~\cite{MS2}, and then experimentally~\cite{Braden1}, that the
leading instability occurs in an antiferromagnetic, not ferromagnetic channel,
and thus a spin-fluctuation exchange in the Berk-Schrieffer spirit would
normally lead to a $d$-wave, not $p$-wave superconductivity.

The issue seems to have been decided conclusively when the Knight shift on Ru
was shown to be temperature-independent across $T_{c}$~\cite{K}, and later
also on O~\cite{K-O}, and the neutron-measured spin-susceptibility was found
to be roughly constant across the transition as well~\cite{chi}. The chiral
$p$-wave state with an order parameter $\mathbf{d}=const(x+iy)\mathbf{\hat
{z},}$ where the Cooper pair spins can freely rotate in-plane, is the only
state that could have this property. Moreover, since in this state spins are
confined in the $xy$ plane, the Knight shift in a magnetic field parallel to
\textbf{\^{z}} is supposed to drop below $T_{c}$ in pretty much the same
manner as in singlet superconductors. Nonetheless, when eventually this
experiment was performed~\cite{Kz}, it appeared that $K_{z}$ is also
independent of temperature. The authors of Ref. \cite{Kz} attempted to
reconcile the accepted pairing symmetry with their experiment, by assuming
that the experimental magnetic field of 200 Oe is affecting drastically the
pairing state and converting it to $\mathbf{d}=f(x,y)\mathbf{\hat
{y}}$ (or the corresponding $x\leftrightarrow y$ partner state).
 One goal of our paper is to estimate whether this
hypothesis is tenable with realistic material parameters.

It is worth noting that the invariance of the in-plane susceptibility is the
only experiment consistent \textit{exclusively} with a chiral $p$-state
(C$p$S). Some probes indicate chirality ($\mu SR$ detects spontaneous currents
below $T_{c}$~\cite{msr}), while others indicate breaking of time-reversal
symmetry~\cite{Kapitulnik}, but the triplet parity is not, in principle,
necessary to explain these experiments. For instance, the singlet chiral state
$\Delta=const(xz+iyz),$ or even $\Delta=const(x^{2}+y^{2}+i\alpha xy),$ which
is not chiral ({although this second state would require two phase transitions
with decreasing temperature, which has never been detected}), but breaks the
time reversal symmetry, are other admissible candidates. Josephson junction
experiments~\cite{Liu} suggested that the order parameter changes sign under
the $(x,y)\leftrightarrow(-x,-y)$ transformation, which is consistent with a
C$p$S, but also with other order parameters~\cite{Zutic}.

{Spin-orbit coupling plays an important role not only in selecting between
different triplet states (chiral vs. planar), but also in the structure of the
chiral state itself. For instance, in Cu$_{x}$Bi$_{2}$Se$_{3},$ instead of the
expected chiral state, a nematic spin-triplet state was observed
~\cite{mat16,yon16}. Indeed, in Cu$_{x}$Bi$_{2}$Se$_{3}$ the large spin-orbit
coupling necessarily implies that a $\mathbf{d}=const(x+iy)\mathbf{\hat{z}}$
induces also an in-plane $\mathbf{d}$-vector component $const(\mathbf{\hat{x}%
}+i\mathbf{\hat{y})}z$~\cite{yip13}. This in-plane component leads to a
non-unitary pairing state, which is not energetically favored in weak
coupling~\cite{fu14}, and, instead, the lower-symmetry nematic state with
$\mathbf{d}=c_{1}x\mathbf{\hat{z}}+c_{2}z\mathbf{\hat{x}}$ is realized. In
principle, similar physics must occur in Sr$_{2}$RuO$_{4}$, but there the
corresponding induced in-plane $\mathbf{d}$-vector component should be much
smaller, thus allowing for a chiral $p$-wave state to exist. However, this is
a quantitative, not qualitative difference, and needs a better understanding
of the role of spin-orbit coupling.}

Finally, recent years have brought about an array of experiments that are
actually \textit{inconsistent} with the C$p$S. { One prediction of a C$p$S is
the existence of edge states at boundaries and at domain
walls~\cite{mat99,kal09,sca15}. However, no evidence for these edge states has
been found~\cite{kir07,kal09}.} There are a variety of predictions about the
response of C$p$S to in-plane magnetic fields that have not been observed
experimentally. In particular, it is known that a finite in-plane magnetic
field should lead to two superconducting transitions as temperature is
reduced~\cite{agt98,VPM} and that the slope of the upper critical field with
temperature at $T_{c}$ should depend upon the in-plane field direction (this
is only true for pairing states that can break time-reversal symmetry)
~\cite{gor84,VPM}. In addition, several different probes indicate behavior
resembling substantial Pauli paramagnetic effects (see Ref.~\cite{PPE} for
discussion and original references).
The latest cloud on the C$p$S sky appeared because of the uniaxial strain
experiments. For the C$p$S (or, in fact, any other two-component state) the
critical temperature, $T_{c},$ under an orthorhombic stress must change
linearly with the strain (the $x\mathbf{\hat{z}}$ and $y\mathbf{\hat{z}}$
state are not degenerate any more, and the splitting is linear in strain). In
the experiment~\cite{strain} $T_{c}$ varies at least quadratically (more
likely, quartically), whereas the linear term is absent within the
experimental accuracy, and only one, very well expressed specific heat jump,
$\Delta C,$ has been observed, with no trace of a second transition even while
the critical temperature changes a lot\cite{hicksAPS}. Moreover, it was
established that both $T_{c}$ and the $\Delta C$ variations trace the changes
in the density of states, and peak when the Fermi level passes the van Hove
singularities at the $X$ or $Y$ points. This observation is particularly
important, because, by symmetry, the superconducting gap in a triplet channel
in a tetragonal superconductor is identically zero at $X$ and $Y$ (it need not
be zero at a finite $k_{z},$ but in a highly 2D material like Sr$_{2}$%
RuO$_{4}$ it will be still very small by virtue of continuity).
Correspondingly, one expects these van Hove singularities to have little
effect on superconductivity. A slightly more subtle, but even more convincing
argument against triplet pairing in Ref.~\cite{strain} is related to the
reduced critical field anisotropy. Finally, a recent detailed study of thermal
conductivity has concluded that a $d$-wave state is by far better consistent
with the thermal transport than the C$p$S~\cite{taillfer}.

In fact, only one fact unambiguously points toward the C$p$S: the invariance
of the spin susceptibility in the in-plane magnetic field --- but, as
discussed above, the analogous experiment for the out-of-plane field $also$
show such an invariance. Thus, our acceptance of the NMR data as an ultimate
proof of the C$p$S hinges upon the possibility of a magnetic field $B\approx
$200 Oe (0.02 T, or 13 mK in temperature units) to overcome the energy
difference between the helical ($\mathbf{d\bot\hat{z})}$ and chiral
($\mathbf{d||\hat{z})}$ states. On can show (the derivation is presented
below) that this implies that the two states, whose energy difference comes
from the spin-orbit (SO) interaction, are nearly degenerate with the accuracy
$\delta\approx10^{-7}$ K$\approx10^{-10}\lambda,$ where $\lambda\approx100$
meV is the SO constant. Moreover, it is often claimed that the solution of
other paradoxes outlined above may be obtained (although nobody has
convincingly succeeded in that) in a formalism where the relativistic effects
would be fully accounted for since the separation between singlet and triplet
channels is only possible in terms of the full angular moment, rather than
just electron spins.

\section{Results and Discussion}

In order to illustrate how SO coupling affects the core assumption of the
field-induced \textbf{d}-vector rotation, let us show a simple
back-of-the-envelope calculation: suppose that the one-electron Hamiltonian
has a relativistic term of the order of $\kappa M_{z}^{2}.$ The physical
meaning of this term is that in the $normal$ state when $n$ electron spins are
confined in the $xy$ plane (as opposed to be parallel to $z$), this affects
the exchange part of the effective crystal potential, and, correspondingly,
one-electron energies. The change is proportional to $n,$ and so is the number
of affected one-electron states, leading to an energy loss of the order of
$\kappa n^{2},$ where $\kappa$ is the magnetic anisotropy scale which is
determined by the SO coupling. One way in which this energy contribution
manifests itself is the conventional magnetic anisotropy in a spin-ordered
state in which case $n\approx M/\mu_{B}.$ However, the same \textquotedblleft
feedback\textquotedblright\ effect must be present in a triplet
superconducting state. The number of electrons bound in Cooper pairs and thus
forced to be either parallel or perpendicular to $z$ can be estimated as
$n\sim\Delta N,$ where $\Delta$ is some average superconducting gap, and $N$
is the density of states, which has been experimentally measured to be about
$8$ states/spin/Ru/eV~\cite{review}. Assuming $\Delta\sim7.5$ K, we estimate
$n\sim0.005$ e/Ru. If the magnetic anisotropy scale $\kappa$ is of the order of
10 K (we will show later that this is the case), then the total energy loss
incurred by rotating the spins of the Cooper pairs is $\Delta E_{sc}%
\approx 2\times10^{-4}$ K (this is smaller than various model estimates of the change
in $T_{c},$ as reviewed in Ref. \cite{MaenoJPSJ}; we use the above estimate
because we wanted to have a conservative $lower$ bound on $\Delta E_{sc}$ and
a \textit{model-independent} estimate of the energy, and not simply a critical
temperature difference, since the latter may, in principle, dramatically
differ from the former). This seems like a small number, but we shall compare
it with the energy gained by allowing screening of an external field of 200 Oe
by Cooper pairs, which is $\Delta E_{mag}\approx \mu_{B}^{2}B^{2}N\approx10^{-7}$ K. This
is \textit{four orders of magnitude} smaller than the estimated loss of
superconducting energy. In other words, to allow for the presumed $d$-vector
rotation, various relativistic effects must fortuitously cancel each other
with a $10^{-3}$ accuracy. Note that in Ref. \cite{MaenoJPSJ}, instead,
$\Delta T_{c}$ was compared with the Zeeman splitting, $\mu_{B}B,$ but this
comparison is hardly relevant at all for the problem at hands; the correct way
is to compare the energy gain with the energy loss.

This simple estimate emphasizes the importance of getting a handle of the type
and scale of relativistic effects in Sr$_{2}$RuO$_{4}.$ So far all efforts in
this direction have been performed either within simplified models or by
educated guesses from the experiment~\cite{ng00,eremin02,annett08,cobo16,scaffidi14}.
The goal of this paper is to
address the issue from a first principle perspective. It is known that this
approach correctly describes (only slightly underestimating) the spin-orbit
interactions\cite{Veenstra} (our SO splitting is exactly the same as
calculated in that reference, 90 meV), and, by comparing the Fourier transform
of the calculated exchange interaction with the experimentally measured
\textbf{q}-dependent spin susceptibility, we observe that the latter is also
well reproduced. The only serious problem with this approach is that it
overestimates the tendency to magnetic ordering for a given set of magnetic
interactions because of the mean field nature of the density functional
theory. Thus we start with a realistic paramagnetic state of Sr$_{2}$%
RuO$_{4},$ using the alloy analogy model in the first principles DFT framework
and calculate the isotropic exchange interactions (See Methods). The Fourier
transform of these interactions gives us the shape of the full spin
susceptibility in the momentum space; as expected, this is peaked at the
nesting vector $\mathbf{q}_{3}=(1,1,0)\frac{2\pi}{3a},$ in agreement with the
experiment. Next, we calculate the mean-field energy of several ordered
magnetic states, all characterized by the same wave vector \textbf{q=q}$_{3}$,
and degenerate without SO interaction. This shall allow us to calculate
 nearest neighbor relativistic Ising terms (see below).
Finally, we calculate magnetic anisotropy for the $\mathbf{q=}(1,0,0)\frac
{\pi}{a}$ states, which breaks the tetragonal symmetry, and from there we
extract the nearest neighbor compass exchange {(See Methods).} The energy
scale of magnetic anisotropy appears rather large, which not only renders the
hypothesis of a \textbf{d}-vector rotation unlikely, but also supports the
idea that anisotropic interactions must be properly accounted for before
drawing conclusions from the experiment. The set of interactions that we
derived should serve as a launching pad and testbed for model calculation of
the superconducting properties. We maintain that a model where all
interelectron interactions are absorbed into spin-spin interactions (with
Hund's coupling between the spins and noninteracting electrons) is
complementary to the widely used Hubbard model and at least as realistic.

Experimentally, Sr$_{2}$RuO$_{4}$ shows no sign of magnetic ordering down to
the low temperatures. However, neutron diffraction studies have
revealed~\cite{sidis99,servant02,braden04} spin-fluctuations in the paramagnetic state with a
characteristic nearly-commensurate wave vector $\mathbf{q}=(0.3,0.3,0)\frac
{2\pi}{a},$ close to $\mathbf{q}_{3}=(1,1,0)\frac{2\pi}{3a},$ which persist
even at the room temperature~\cite{Iida}. The Density Functional Theory (DFT),
being a static mean field theory (by some criteria, the best such theory
possible), overestimates the tendency to magnetism. In its Generalized
Gradient Approximation flavor (GGA) DFT stabilizes even ferromagnetic order,
albeit with small moments~\cite{deGroot}. Unsurprisingly, spin density waves
with $\mathbf{q=q}_{3}$ are even lower in energy. This deficiency of the DFT
can, however, be put to a good use by mapping the DFT (\textit{i.e.,
}mean-field) energetics onto a spin-Hamiltonian, as it is often done, for
instance, for Fe-based superconductors~\cite{Glasbrenner}.
Since the isotropic and anisotropic magnetic interactions entails completely
different energy scales, and require different level of accuracy, we have chosen
two different techniques to calculate them; as discussed below, the isotropic
calculations were performed perturbatively, allowing us to fully account for
the long-range, nesting-driven interaction, while the nearest neighbor exchange
interactions were calculated by brute force comparing highly accurate energy
values in different magnetic configurations.

First we have calculated the Heisenberg part of the Hamiltonian, defined as%
\begin{equation}
H_{H}=-\sum_{<i\neq j>}J_{ij}\mathbf{M}_{i}\mathbf{\cdot M}_{j}%
,\label{heisenberg}%
\end{equation}
where $\mathbf{M}_{i}$ is the Ru moment on the site $i,$ and the summation is
performed over all bonds up to a given coordination sphere. The parameters are
calculated in the Disordered Local Moments (DLM) approximation~\cite{Gyorffy},
which is used to model the paramagnetic state of Sr$_{2}$RuO$_{4}$ {(See
Methods for more details and employed approximations)}.

\begin{figure}[t]
\begin{center}
\includegraphics[width=85mm]{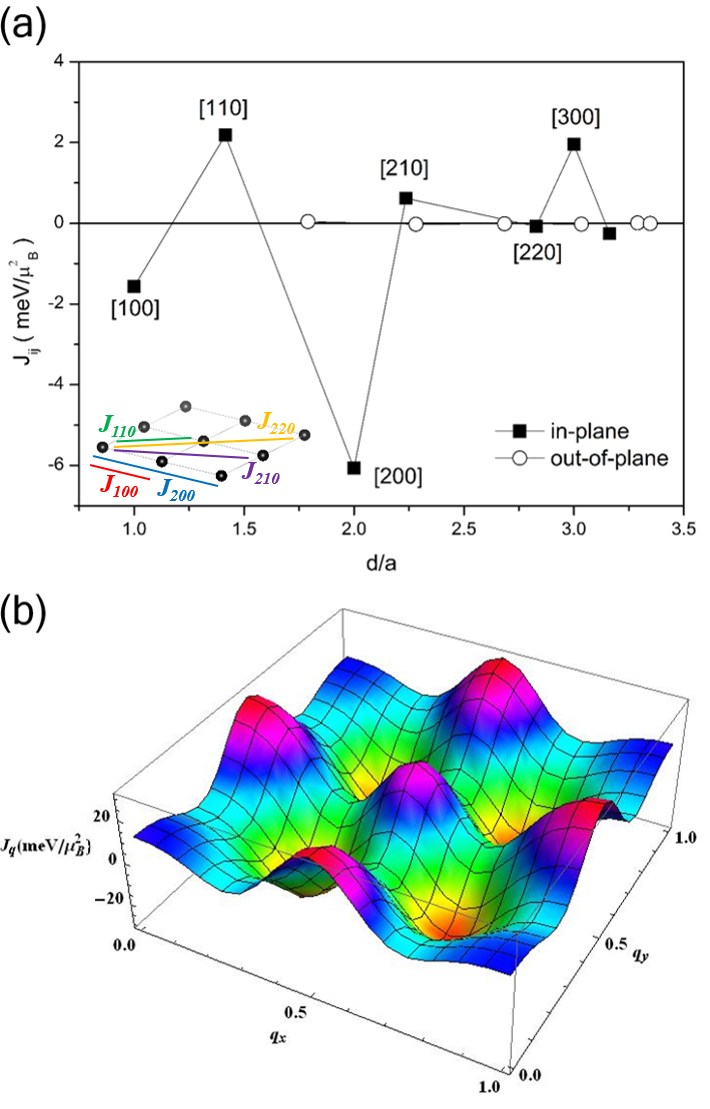}
\end{center}
\caption{\textbf{Calculated exchange interactions up to the 7th coordination
sphere in Sr$_{2}$RuO$_{4}$.} (a) The distance dependence (in terms of planar
lattice constant) of isotropic exchange interactions for in-plane (filled
square) and out-of-plane (open circle). (b) The Fourier transform of the
exchange interactions shown in the panel (a).}%
\label{kkr}%
\end{figure}

The results presented in the Fig.~\ref{kkr} are derived for the Ru local
moment being fixed to 1 $\mu_{B}$ in the DLM state. The obtained values of the
exchange constants, however, are fairly independent of the values of the local
moment fixed in the DLM state; the minimum of the Fourier transform is always
at $\mathbf{q}=(\alpha,\alpha,0)$ with $\alpha=0.3-0.31$. Note that the
interplane exchanges nearly vanish, indicating an almost perfect 2D character
of the magnetism in Sr$_{2}$RuO$_{4}$. For instance, the nearest neighbors
between-the-planes $J_{001}\approx0.5-1$ K/$\mu_{B}^{2}$ (ferromagnetic), or
about $-0.01J_{200}.$

The leading term is the in-plane third nearest neighbour (NN)
antiferromagnetic interaction $J_{200}$, which is quite counterintuitive from
the point of view of the Hubbard model and superexchange mechanism that is
often employed as a starting point. This is a consequence of the Ru electrons
itinerancy, since Sr$_{2}$RuO$_{4}$ is a metal. The lattice Fourier transform,
$J(\mathbf{q})$, of the calculated interactions is shown in the
Figure~\ref{kkr} (b). $J(\mathbf{q})$ has a meaning of a measure of the energy
($J(\mathbf{q})\cdot M^{2}$) of the spin-density fluctuations with a
wave-vector $\mathbf{q}$ and a given amplitude $M$ [the quantitity that is
directly related to the static zero-temperature spin susceptibility is
$1/J(\mathbf{q})]$ The deep minima of $J(\mathbf{q})$ at $\mathbf{q}%
=(0.31,0.31,0)\frac{2\pi}{a}$ suggest that the spin-fluctuations with the wave
vector $\mathbf{q}$ will be dominant in the paramagnetic state of Sr$_{2}%
$RuO$_{4}$. The position of these minima is indeed in perfect agreement with
the sharp maxima of the integrated magnetic scattering intensity,
experimentally observed in neutron diffraction~\cite{Iida}. Thus, both our
calculation and the experiment suggest the dominance of the spin-fluctuations
with the wave vector $\mathbf{q}_{3}$ in the excitation spectra of Sr$_{2}%
$RuO$_{4}$.

In order to extract the relevant anisotropic exchange interaction parameters,
we used direct calculations of the total energy in different magnetic
configurations compatible with the ordering vector $\mathbf{q}_{3}$. Note that
anisotropic magnetic interactions appear exclusively due to the SO coupling
{(See Methods for the description of codes and approximations used in these
calculations)}. Allowed anisotropic terms for the nearest neighbor terms are
absorbed in the following Hamiltonian (simplified compared to a more complete
expression discussed in the Methods section):%
\begin{align}
H_{rH} &  =H_{H}+\sum_{<nn>}J^{zz}M_{i}^{z}M_{j}^{z}\nonumber\\
&  +\sum_{<nnx>}J^{xy}(M_{i}^{x}M_{j}^{x}-M_{i}^{y}M_{j}^{y})\nonumber\\
&  +\sum_{<nny>}J^{xy}(M_{i}^{y}M_{j}^{y}-M_{i}^{x}M_{j}^{x}%
),\label{anisotropic}%
\end{align}
where the first term is given by Eq. \ref{heisenberg}, the second is Ising
exchange (sometimes called the Kitaev interaction), and the last two represent
the compass term. Summation in the last two terms is over all horizontal and
all vertical bonds, respectively, while in the Ising term it is over all
inequivalent bonds. Note that Dzyaloshinskii-Moriya terms~\cite{Dzyaloshinskii,Moriya} are not
allowed by symmetry.

\begin{figure}[t]
\begin{center}
\includegraphics[width=85mm]{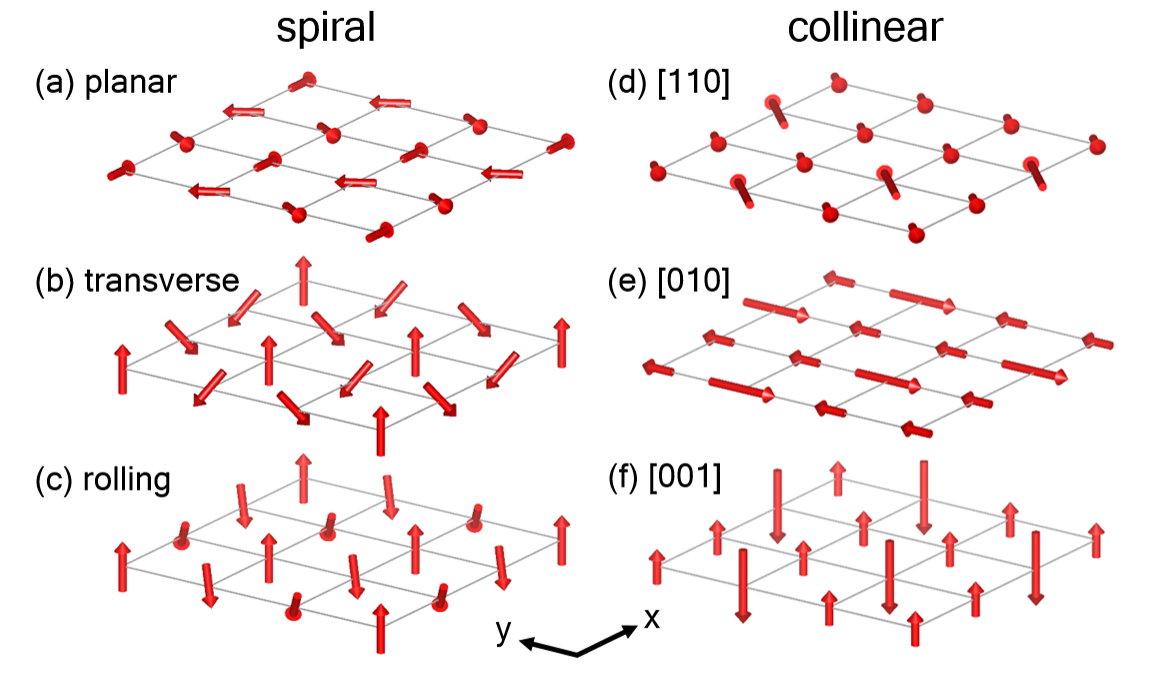}
\end{center}
\caption{ \textbf{Lowest energy magnetic structures ($\mathbf{q}
=(1,1,0)\frac{2\pi}{3a}$) of RuO$_{2}$ basal plane in Sr$_{2}$RuO$_{4}$.} The
(a)-(c) structures represent different types of spiral magnetic order and
(d)-(f) corresponds to the collinear up-up-down magnetic order with different
moment directions.}%
\label{mag_str}%
\end{figure}

The six most energetically favorable states are depicted in Fig.~\ref{mag_str}%
. The first three states can be described as harmonic spin-density waves
(SDWs):%
\begin{equation}
\mathbf{M}_{ijk}=m\mathbf{A}\exp(-i\mathbf{R}_{ijk}\cdot\mathbf{q}%
_{3}),\label{spirals}%
\end{equation}
where $\mathbf{A}_{a}=[-\frac{1}{2},\frac{\sqrt{3}}{2},0],$ $\mathbf{A}%
_{b}=[\frac{i}{2\sqrt{2}},-\frac{i}{2\sqrt{2}},\frac{1}{2}],$ $\mathbf{A}%
_{c}=[\frac{i}{2\sqrt{2}},\frac{i}{2\sqrt{2}},\frac{1}{2}],$ with $m$ hardly
varying between the three states and equal to 0.76 $\mu_{B}.$ The fourth to
sixth states are collinear where the amplitude of the moments varies along
each of the crystallographic directions 100, 010, and 110 as $m^{\prime
},-m^{\prime}/2,-m^{\prime}/2$ (more precisely, $1.07,-0.56,-0.56$ $\mu_{B}).$
Note that $m^{\prime}$ is very close to $\sqrt{2}m$ in the harmonic SDWs, and
the average $\left\langle \mathbf{M}^{2}\right\rangle $ is the same in all
these states (within a 1.3\% error). In this collinear state the direction of
the magnetization can be selected in three inequivalent ways, namely along
110, 1\={1}0 or 001. Upon inclusion of the SO term, the 001 collinear
up-up-down structure is the ground state (Table~\ref{table1}).

Next, we fit the energy differences in Table~\ref{table1} to the Hamiltonian
(\ref{anisotropic}), extracting $J^{zz}$ and $J^{xy}$ (the fitting procedure
included more parameters than in Eq. \ref{anisotropic}, and is discussed in
the Methods Section). All isotropic (Heisenberg) parts of the exchange
interactions are included in the $H_{H}$. The compass parameter $J^{xy}$ does
not affect the states with $\mathbf{q}\varpropto(1,1,0),$ and was extracted
from a separate set of calculations with $\mathbf{q}_{2}=(0,1,0)\frac{\pi}%
{a},$ and $\mathbf{M}_{ijk}=m\mathbf{A}\exp(-i\mathbf{R}_{ijk}\cdot
\mathbf{q}_{2}),$ where $\mathbf{A}_{\perp}=(1,0,0)$ and $\mathbf{A}%
_{||}=(0,1,0),$ and $m$ was fixed to be equal to its value in the spiral
states, 0.76 $\mu_{B}$. These wave vectors define so-called single stripe
antiferromagnetic order, well known in Fe-based superconductors.

\begin{table}[h]
\caption{Calculated total energies (meV/Ru) of various states with the
$\mathbf{q}_{3}=(1,1,0)\frac{2\pi}{3a}$ periodicity. For spiral phases, the
magnitude of the calculated local moments is 0.76$\mu_{B}$, and for collinear
up-up-down phase, 0.57$\mu_{B}$ and 1.03$\mu_{S}$ for up and down spin,
respectively. }%
\label{table1}%
\centering
\begin{ruledtabular}
\begin{tabular}
[c]{l|c|c|c}
${\bf q}$&& spin orientation&energy\\
\hline
&& planar & 0 \\
$(1,1,0)(2\pi/3a)$&{spiral} & rolling & -0.42 \\
&&transverse& -0.22 \\\hline
&& (110) & -0.34 \\
$(1,1,0)(2\pi/3a)$&collinear& (010) & -0.24  \\
&& (001) & -1.27 \\
\hline
$(1,0,0)(\pi/a)$&collinear&$(100)$&38.06\\
&stripes&$(010)$&39.57
\\
\end{tabular}
\end{ruledtabular}  \end{table}

Thus obtained parameters are $J^{zz}=-1.2\pm0.6$ meV/$\mu_{B}^{2}$, and
$J^{xy}=1.0$ meV/$\mu_{B}^{2}$ ($J^{zz}m^{2}=-0.70\pm0.35$ meV, $J^{xy}=0.57$
meV, for $m=0.76$ $\mu_{B}$). The details of the fitting are described in the
Methods section. Note that $J^{xy}$ does not have an error bar not because it
was accurately determined, but because we did not have enough calculations to
estimate the error. First, one observes that the scale of the anisotropy
induced by SO is of the order of $10$ K. As discussed in the introduction,
this renders the explanation of the invariance of the Knight shift below
$T_{c}$ in term of the order parameter rotation~\cite{Kz} untenable and shakes
the main argument in favor of the chiral triplet superconductivity in Sr$_{2}%
$RuO$_{4}$. Second, our fitting provides a powerful tool for modeling normal
and especially superconducting properties of Sr$_{2}$RuO$_{4}$ from an
entirely different perspective. Compared to the generally accepted models
based on the Hubbard-Hund Hamiltonians, our new approach is based entirely on
first principles calculations, and emphasizes the role of magnetic
interactions. The corresponding \textit{DFT-inspired }model Hamiltonian reads%
\begin{align}
H &  =H_{rH}+H_{e}\label{DEX0}\\
H_{e} &  =\sum_{\mathbf{k}\alpha s}\varepsilon_{\mathbf{k}\alpha}%
c_{\mathbf{k}\alpha s}^{\dagger}c_{\mathbf{k}\alpha s}-I\sum_{\mathbf{kq}%
\alpha ss^{\prime}}c_{\mathbf{k-q,}\alpha s}^{\dagger}\mathbf{M}%
_{q}\mathbf{\cdot\sigma}_{ss^{\prime}}c_{\mathbf{k}\alpha s^{\prime}%
},\label{DEX}%
\end{align}
where the first term is the noninteracting energy, with the band (spin)
indices $\alpha$ ($s$), and the second is the Hund's rule (Stoner, in the DFT
parlance) coupling. All electron-electron interactions carried by spin
fluctuations are absorbed in the local Hund's interaction $I$ and the
intersite magnetic interactions $H_{rH},$ while interactions due to charge
fluctuations are not included in Eq.~\ref{DEX}, but can be added separately,
if needed (or just collected in one Coulomb pseudopotential $\mu^{\ast},$ as
in the Eliashberg theory). Eq.~\ref{DEX} can be understood as a generalized
double-exchange Hamiltonian~\cite{Khomskii}. Indeed, this model, inspired by
DFT calculations, entails electrons moving in the same effective potential as
used in other techniques, and described by the same tight-binding parameters.
However, as it is usual in DFT, all electron-electron interactions are
implicitly integrated out. Instead, we introduce quasi-local magnetic moments
that interact with the electrons via the local Hund's rule coupling
(parameterized as the Stoner parameter in DFT), while the moments interact
among themselves according to the sum of the long-range Heisenberg and the
short-range anisotropic Hamiltonian (Eq.~\ref{anisotropic}). The former part
incorporates implicitly all Fermi surface effects, including nesting at
$\mathbf{q}=\{0.3,0.3,0\}\frac{2\pi}{a},$ while the latter selects between
different triplet states. It is important not to attempt to integrate out the
free carriers $c_{\mathbf{k}\alpha s}$ in Eq. \ref{DEX} in order to extract
additional interaction between the local moments \textbf{M;} that would have
been incorrect, because all such interactions had been computed previously and
embedded in $H_{rH}.$ \textbf{ }On the contrary, the intended solution of
these equations is integrating out the \textbf{M}'s in order to obtain the
effective pairing interaction, as illustrated below.

It might be instructive to demonstrate how Eqs. \ref{DEX0}, \ref{DEX} can be
reduced to a Hamiltonian including only the itinerant electrons (as convenient
for analyzing superconductivity). We can safely assume that all $J$s are much
smaller than $I,$ introduce the itinerant spin polarization $\mathbf{s}%
_{i\alpha}=\sum_{ss^{\prime}}c_{i\alpha s}^{\dagger}\mathbf{\sigma
}_{ss^{\prime}}c_{i\alpha s^{\prime}},$ and single out the terms relevant to
the pairwise interaction between $\mathbf{s}_{i\alpha}$ and $\mathbf{s}%
_{j\beta}:$%
\begin{equation}
E_{ij,\alpha\beta}=-I\mathbf{M}_{i}\cdot\mathbf{s}_{i\alpha}-I\mathbf{M}%
_{j}\cdot\mathbf{s}_{i\beta}-J_{ij}\mathbf{M}_{i}\cdot\mathbf{M}_{i}%
\end{equation}
In the lowest order in $J,$ the mean field solution requires that
$\mathbf{M}_{i}$ and $\mathbf{s}_{i\alpha}$ be parallel, $E_{ij,\alpha\beta
}=-2IMs-J_{ij}M^{2}\mathbf{\hat{s}}_{i\alpha}\cdot\mathbf{\hat{s}}_{i\beta},$
and the effective pairwise interaction can be written as $-J_{ij}%
M^{2}\mathbf{\hat{s}}_{i\alpha}\cdot\mathbf{\hat{s}}_{i\beta}$ (note that
essentially the same Hamiltonian, only written in the orbital basis rather
than the band basis, which can also be done in this case, was applied to
Fe-based superconductors in several papers, for instance, in Ref.~\cite{Hu};
after summation of the total energy over the band indices $\alpha,\beta$ these
approaches become equivalent). In principle, one can easily derive the next
order correction to the interaction, which is $+(J_{ij}^{2}M^{3}%
/Is)(\mathbf{\hat{s}}_{i\alpha}\cdot\mathbf{\hat{s}}_{i\beta})^{2}.$

As an example of how this Hamiltonian can be used to address
superconductivity, we solve in the simplest mean field approximation the
problem of the relative energetics of the five unitary $p-$triplet states. In
particular, beginning with $H=-J_{ij}M^{2}\mathbf{\hat{s}}_{i\alpha}%
\cdot\mathbf{\hat{s}}_{i\beta}$ and restricting the electronic spins to a
single band for simplicity (generalizing onto three bands with realistic
dispersions is straighforward), we find that the Ising and compass exchange
modify the pairing interaction $\delta V$ in different pairing channels
differently, as shown in Table \ref{table3}.

\begin{table}[ptb]
\caption{Relative change in pairing interaction for spin-triplet pairing
channels due to Ising and compass exchange terms}%
\label{table3}
\begin {ruledtabular}
\begin{tabular}{|c|c|c|}
Pairing Channel & &$\delta V/2$ \\
\hline
$(\sin k_x\pm i\sin k_y)\hat{z}$&axial chiral & $J^{zz}$  \\
$\sin k_x \hat{x}+\sin k_y \hat{y}$&planar radial& $-J^{zz}+2J^{xy}$ \\
$\sin k_x \hat{y}+\sin k_y \hat{x}$&planar quadrupolar&  $-J^{zz}-2J^{xy}$  \\
$\sin k_x \hat{x}-\sin k_y \hat{y}$&planar quadrupolar & $-J^{zz}+2J^{xy}$ \\
$\sin k_x \hat{y}-\sin k_y \hat{x}$&planar tangential & $-J^{zz}-2J^{xy}$ \\
\end{tabular}
\end {ruledtabular}
\end{table}

Thus, in this approximation the five states split into two planar doublets (of
course, this degeneracy is not driven by symmetry, and will be lifted in more
sophisticated calculations, but likely the splitting will be small) and a
C$p$S singlet, which is located between the doublets if $J_{zz}>-|J_{xy}|$ and
below both of them otherwise (note that we found $J_{zz}$ to be negative. In other words, we have shown that selection
between chiral and planar superconductivity is driven by the competition
between the Ising and compass anisotropic exchange. Of course, this is just an
illustration of principle; in principle, this approach should be applied to
the true three-band electronic structure and extended to singlet as well as
triplet states, but this is beyond the scope of this paper.

We reiterate that we do not insist that this approach is \textit{superior} to
the Hubbard Hamiltonian, but it is \textit{different} and
\textit{complementary,} having the potential to uncover new physics. Similar
to the former, it can be used in the contexts of, $e.g.,$ random phase
approximation (RPA), fluctuation exchange (FLEX) or functional renormalization
group (fRG) calculations.

\begin{figure}[t]
\begin{center}
\includegraphics[width=85mm]{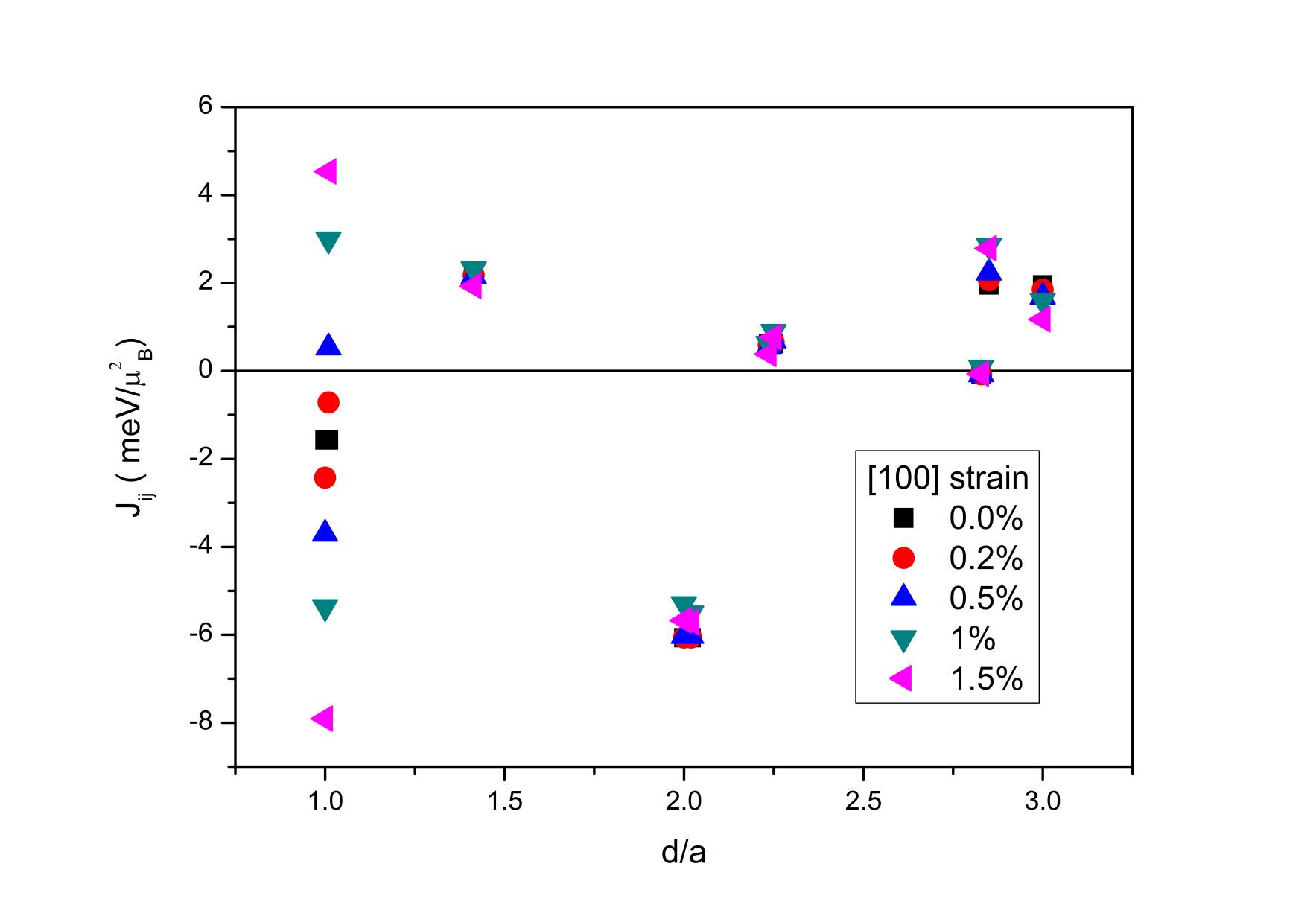}
\end{center}
\caption{\textbf{Same as Fig. \ref{kkr}a, but as a function of uniaxial
strain.} Only the nearest neigbor exchange constant is affected by the strain
(split into $J_{a}$ and $J_{b}$) at a noticeable level.}%
\label{strain}%
\end{figure}

A final note relates to the recent experiments on strained Sr$_{2}$RuO$_{4}.$
This is a large topic mostly outside of the scope of this paper. However, we
would like to make one comment in this regard. The fact that $T_{c}$ rapidly
grows with the strain and peaks at the strain corresponding to the Lifshits
transition (where the $\gamma$ band touches the X-point) can be explained by
either a DOS effect (van Hove singularity) or by a change in pairing
interaction. The former explanation, as mentioned before, is realistic for
singlet, but not triplet pairing symmetries. The latter would be viable if the
change in DOS were sufficient to shift the balance between the AF and FM
tendencies toward the latter. To verify that, we have repeated the
calculations of the Heisenberg parameters in the strained case. However, we
found that the main effect of the strain is not related to the van Hove
singularity, and that the average exchange coupling does not become more
ferromagnetic. Instead, the strain introduces a splitting between $J_{1a}$ and
$J_{1b},$ while the average value barely changes, as shown in
Fig. \ref{strain}. These results therefore indicate that the peak in $T_{c}$ is directly related
to the peak in DOS, and not $via$ enhanced pairing interaction.
This conclusion is supported by recently reported
thermodynamic results\cite{hicksAPS}, which strongly suggest that not only
$T_{c},$ but also $\Delta C/T_{c}$ is peaked at the van Hove singularity.

To summarize, we have presented first principle calculations of the leading
isotropic and anisotropic magnetic interactions in Sr$_{2}$RuO$_{4}$. Our
results indicate that rotating a $p-$wave superconducting order parameter
during measurements of the Knight shift is impossible by several orders of
magnitude and thus the invariance of the Knight shift across the transition
remains an unresolved puzzle. We further proposed a model framework, based on
a double-exchange type Hamiltonian, and incorporating the calculated magnetic
interactions in their entirety, and present an example of using this framework
for addressing superconducting pairing symmetry.


\section{Methods}


\textbf{First principles calculations}

For relativistic total energy calculations we have employed the projector
augmented wave method~\cite{PAW} as implemented in the Vienna Ab initio
Simulation Package (VASP)~\cite{Kresse}, including SO coupling~\cite{SOCVASP}.
We have used the density functional theory within the Perdew-Burke-Ernzerhof
parametrization for the exchange and correlation potential~\cite{PBE}, and the
experimental lattice structure is employed in all calculations. The energy
cutoff was set to 400 eV with convergence criteria of $10^{-6}$~eV. We used up
to 1386 irreducible k-points, reduced to 900 for the four formula units cell.
For Ru, a pseudopotential with $p-$states includes as valence states was selected.

For the calculation of the isotropic exchange constants we used the
Korringa-Kohn-Rostokker KKR method within the atomic sphere approximation
(ASA)~\cite{KKRASA} and the Green function based magnetic-force
theorem~\cite{Lichtenstein}. The implementation of this technique has been
described elsewhere~\cite{MFT-KKR}. Physically, this technique can be
considered to be a magnetic analogue of the disordered alloys theory based on
coherent potential approximation~\cite{MFT-KKR} and is known as the Disordered
Local Moments (DLM) approximation~\cite{Cyrot,Gyorffy}. Upon fixing the Ru
magnetic moments in the DLM state we achieved self-consistency using 115
irreducible k-points in the Brillouin zone, and then used an extended set of
k-points (1529) to compute the isotropic exchange constants in the framework
of the magnetic force theorem.

\textbf{Fitting procedure. }The full equation used to describe the calculated
energies, including all bilinear terms up to the second neighbors, reads:

\begin{widetext}\begin{eqnarray}
H_{r}  &  =&\sum_{i}K(M_{i}^{z})^{2}+\sum_{<nn>}J_{1}^{zz}M_{i}^{z}M_{j}^{z}\nonumber\\
&  +&\sum_{<100>}J_{1}^{xy}(M_{i}^{x}M_{j}^{x}-M_{i}^{y}M_{j}^{y})+\sum_{<010>}J_{1}^{xy}(M_{i}^{y}M_{j}^{y}-M_{i}^{x}M_{j}^{x})\label{ani}\\
&  + & \sum_{<nnn>}J_{2}^{zz}M_{i}^{z}M_{j}^{z} + \sum_{<110>}J_{2}^{xy}(M_{i}^{x}M_{j}^{x}-M_{i}^{y}M_{j}^{y})
+\sum_{<1\bar{1}0>}J_{2}^{xy}(M_{i}^{y}M_{j}^{y}-M_{i}^{x}M_{j}^{x}) \label{last}
\end{eqnarray}
\begin{table}[t]
\caption{Energies of various calculated magnetic states and the corresponding coefficients
in Eq. \protect\ref{ani}. Magnetic moments are described as
$
\mathbf{M}_{ijk}=\operatorname{Re}[m\mathbf{A}\exp(-i\mathbf{R}_{ijk}%
\cdot\mathbf{q}_{3})]$.
}%
\label{table1M}
\centering
\begin{ruledtabular}
\begin{tabular}
{c|c|c|c|c}
& &${\bf A}$ &$m$& energy  \\ \hline
&planar&$\{1,-i,0\}$ & 1 & 0 \\
spiral&rolling&$\{\frac{i}{\sqrt{2}},\frac{i}{\sqrt{2}},1\}$ & 1 & $(K-J{zz})/2+J_2^{zz}/4-3J_2^{xy}/4 $\\
&transverse&$\{\frac{i}{\sqrt{2}},\frac{-i}{\sqrt{2}},1\}$&1& $(K-J{zz})/2+J_2^{zz}/4+3J_2^{xy}/4 $\\\hline
& (110)& $\{\frac{i}{\sqrt{2}},\frac{i}{\sqrt{2}},0\}$ &$\sqrt{2},1/\sqrt{2}$&$-3J_2^{xy}/2 $\\
collinear& (100)&$\{-1,0,0\}$&$\sqrt{2},1/\sqrt{2}$&0 \\
& (001)& $\{0,0,-1\}$&$\sqrt{2},1/\sqrt{2}$&$K-J{zz}+J_2^{zz}/2 $\\
\hline
\end{tabular}
\end{ruledtabular}  \end{table}
\end{widetext}

Here, for completeness, we have included  the single-site anisotropy term $K$;
since it always enters in the same combination with $J^{zz}$, they cannot be
decoupled within this set of calculations. While this term is, in principle,
allowed because of itinerancy, we note that the calculated magnetization is
close to the the $S=1/2$ and therefore we expect $K\ll J^{zz}$. This
approximation was used in the main text. We have also included, besides the
nearest neighbor anisotropic interaction $J_{1}^{zz}$ and $J_{1}^{xy},$ the
corresponding second nearest neighbor interactions $J_{2}^{zz}$ and
$J_{2}^{xy}.$ The latter distinguishes between the collinear state polarized
along the (110) tetragonal direction and the one polarized along (100), and
the transverse and rolling spirals. We found it to be relatively small,
$0.17\pm0.05$ meV/$\mu_{B}^{2}.$ The second nearest neighbour Ising
interaction $J_{2}^{zz}$ simply adds to $J^{zz},$ and therefore was absorbed
into the latter in the fitting procedure. The difference in energies between
the planar spiral and the (100) collinear structure, $0.24$ meV, is likely
related to the fact that the isotropic exchange constants enter these two
state differently. Our non-relativistic calculations find them degenerate
within the computational accuracy, apparently, fortuitously. Since SOC also
affects the isotropic constants, it is no surprise that relativistic effects
break this accidental degeneracy.

One can calculate $K-J^{zz}$ and $J_{2}^{xy}$ either from the set of spiral
calculations, or from collinear calculations; the results differ by $\pm30$\%.
It is unlikely that this is due to computational inaccuracy, but rather to
other interactions not accounted for, such as third neighbors (which is the
leading isotropic exchange) or anisotropic biquadratic coupling.

The full summary of the magnetic patterns and their energies used for the
fitting, as well as the expressions for the total energies in terms of the
parameters in Eq. \ref{last}, are presented in Table \ref{table1M}.

\textbf{Mean-field comparison of pairing energies}. To find the interactions
in Table II, we begin with the following Hamiltonian $H_{int}$ that includes
charge and spin fluctuations. As an example of how this approach can be used
we ask a relatively simple question of how the magnetic anisotropy we have
found affects spin-triplet pairing states. To this end, we generalize Ref.
\cite{miy86} and consider only a single band with the following  Hamiltonian
with charge, $\rho(q),$ and spin, $S_{i}(q),$ interactions
\begin{align*}
H_{int}  & =\sum_{q}\left[  U(q)\rho(q)\rho(-q)+\sum_{i}J_{i}(q)S_{i}%
(q)S_{i}(-q)\right]  \\
& =\sum_{k,k^{\prime}}\sum_{q}a_{k+q/2,s}^{\dagger}a_{-k+q/2,s^{\prime}%
}^{\dagger}a_{-k^{\prime}+q/2,m^{\prime}}a_{k^{\prime}+q/2,m}\\
& \times[\rho(k-k^{\prime})\delta_{s,m}\delta_{s^{\prime},m^{\prime}}%
+J_{z}(k-k^{\prime})\sigma_{s,m}^{z}\sigma_{s^{\prime},m^{\prime}}^{z}\\
& +J_{x}(k-k^{\prime})\sigma_{s,m^{\prime}}^{x}\sigma_{s^{\prime},m}^{x}%
+J_{y}(k-k^{\prime})\sigma_{s,m^{\prime}}^{y}\sigma_{s^{\prime},m}^{y}]
\end{align*}
Focussing on superconductivity with zero momentum Cooper pairs, $H_{int}$ can
be rewritten
\[
H_{int}=\frac{1}{2}\sum_{k,k^{\prime}}\Big [V_{s}(k-k^{\prime})s_{k}^{\dagger
}s_{k^{\prime}}+\sum_{i=x,y,z}V_{t,i}(k-k^{\prime})t_{i,k}^{\dagger
}t_{i,k^{\prime}}\Big]
\]
where $s_{k}=\sum_{s,s^{\prime}}(i\sigma_{y})_{s,s^{\prime}}c_{-k,s}%
c_{k,s^{\prime}}$ and $t_{i,k}=\sum_{s,s^{\prime}}(i\sigma_{i}\sigma
_{y})_{s,s^{\prime}}c_{-k,s}c_{k,s^{\prime}}$ are the possible singlet and
triplet Cooper pair operators, and the effective interactions for the
different pairing channels are found to be
\begin{align*}
V_{s} &  =\rho(k-k^{\prime})-J_{x}(k-k^{\prime})-J_{y}(k-k^{\prime}%
)-J_{z}(k-k^{\prime})\\
V_{t,x} &  =\rho(k-k^{\prime})-J_{x}(k-k^{\prime})+J_{y}(k-k^{\prime}%
)+J_{z}(k-k^{\prime})\\
V_{t,y} &  =\rho(k-k^{\prime})+J_{x}(k-k^{\prime})-J_{y}(k-k^{\prime}%
)+J_{z}(k-k^{\prime})\\
V_{t,z} &  =\rho(k-k^{\prime})+J_{x}(k-k^{\prime})+J_{y}(k-k^{\prime}%
)-J_{z}(k-k^{\prime}).\\
&
\end{align*}
This result reduces to that found when spin interactions are isotropic
\cite{miy86,cho13} or have uniaxial symmetry \cite{kuw00}. In our case, the
specific form of the spin anisotropy is
\begin{align*}
J_{z} &  =2J_{z0}[\cos(k_{x}-k_{x}^{\prime})+\cos(k_{y}-k_{y}^{\prime})]\\
J_{x} &  =2J_{\perp0}[\cos(k_{x}-k_{x}^{\prime})-\cos(k_{y}-k_{y}^{\prime})]\\
J_{y} &  =2J_{\perp0}[\cos(k_{y}-k_{y}^{\prime})-\cos(k_{x}-k_{x}^{\prime})]\\
&
\end{align*}
Expressing $H_{int}$ with the above spin anisotropy in terms of irreducible
representations of tetragonal symmetry for the Cooper pairs leads to Table II.

\section{Acknowledgements}

B.K, S.K, and C.F were supported by the joint FWF and Indian Department of
Science and Technology (DST) project INDOX (I1490-N19), and by the FWF-SFB
ViCoM (Grant No. F41). I.I.M. is supported by ONR through the NRL basic
research program.  D.F.A was supported by the National Science Foundation grant
DMREF-1335215.

\section{Contributions}

S.K. and I.M conceived the research; B.K has carried out most of the numerical
calculations with contributions by S.K. and I.M; the superconductivity-related
discussion was authored by D.A., who also performed the sample mean-field
calculations, and by I.M. All authors participated in the discussions and
contributed to writing the paper; C.F. supervised the Vienna part of the project.

\end{document}